\documentclass[prl,aps,showpacs,floatfix,twocolumn]{revtex4-1}
\usepackage{amsmath}
\usepackage{amsfonts}
\usepackage{amssymb}
\usepackage{revsymb}
\usepackage{graphicx}
\usepackage{physics}
\usepackage{pdfpages}

\makeatletter
\AtBeginDocument{\let\LS@rot\@undefined}
\makeatother

\newcommand{\eref}[1]{Eq.~(\ref{#1})}

\begin{document}
\title{Enhanced micromotion compensation using a phase modulated light field.}
\author{K. J. Arnold}
\affiliation{Centre for Quantum Technologies, National University of Singapore, 3 Science Drive 2, 117543 Singapore}
\author{N. Jayjong}
\affiliation{Centre for Quantum Technologies, National University of Singapore, 3 Science Drive 2, 117543 Singapore}
\author{M. L. D. Kang}
\affiliation{Centre for Quantum Technologies, National University of Singapore, 3 Science Drive 2, 117543 Singapore}
\author{Qin Qichen}
\affiliation{Centre for Quantum Technologies, National University of Singapore, 3 Science Drive 2, 117543 Singapore}
\author{Zhao Zhang}
\affiliation{Centre for Quantum Technologies, National University of Singapore, 3 Science Drive 2, 117543 Singapore}
\author{Qi Zhao}
\affiliation{Centre for Quantum Technologies, National University of Singapore, 3 Science Drive 2, 117543 Singapore}
\author{M. D. Barrett}
\affiliation{Centre for Quantum Technologies, National University of Singapore, 3 Science Drive 2, 117543 Singapore}
\affiliation{Department of Physics, National University of Singapore, 2 Science Drive 3, 117551 Singapore}
\affiliation{National Metrology Center, Agency for Science, Technology and Research (A*STAR), Singapore}
\begin{abstract}
We investigate sideband spectroscopy of a trapped ion using a probe laser phase modulated at the trap drive frequency.  The enhanced sensitivity of our technique over traditional sideband spectroscopy allows us to detect stray fields of $0.01\,\mathrm{V/m}$ on a timescale of a few minutes and detect differential phases of $5\,\mu\mathrm{rad}$ between applied ac potentials.  We also demonstrate the ability suppress Doppler shifts from excess motion to well below the limit imposed by the intrinsic motion of the ion in the vibrational ground-state.  The technique we introduce can be readily implemented in any ion trap system that utilizes sideband spectroscopy for micromotion compensation and can be seamlessly integrated into experiments in a fully automated way.
\end{abstract}
\maketitle
Micromotion in linear Paul traps is the small-amplitude, rapid motion of the ion driven by the radio frequency (rf)  trapping potentials \cite{berkeland1998minimization, keller2015precise}.  It consists of unavoidable intrinsic micromotion (IMM) arising from the thermal motion of the ion and so called excess micromotion (EMM) arising from residual rf electric fields at the trap equilibrium position.  The effects of micromotion are typically detrimental and minimizing EMM is an essential part of ion-trap experiments.  This is particularly true for trapped-ion frequency standards for which micromotion is an important contribution to uncertainty budgets \cite{huntemann2016single,brewer2019al+,zhiqiang2023176lu+}, or for hybrid atom-ion experiments for which micromotion can dominate collision dynamics \cite{meir2018experimental,kalev2012micromotion,harter2013minimization}.

A variety of techniques for EMM detection have been investigated \cite{berkeland1998minimization,keller2015precise,harter2013minimization,gloger2015ion,chuah2013detection,allcock2010implementation} with the choice of technique constrained by the experimental setups.  Photon correlation and resolved sideband methods \cite{berkeland1998minimization, keller2015precise} are widely used due to their general applicability.  Both methods make use of radio-frequency (rf) sidebands at the trap drive frequency $\Omega_\mathrm{rf}$ as seen in the rest frame of the ion.  Photon correlation methods utilize a transition having a linewidth $\Gamma \gtrsim \Omega_\mathrm{rf}$, which is typically the transition used for state sensitive detection and cooling.  In this regime, ion motion effectively modulates the laser detuning and the rate of fluorescence is correlated with the phase of the rf voltage.  The phase relationship is advantageous as it also enables the distinction between EMM due to ion displacement from the rf null and that due to a phase difference between voltages applied to rf electrodes \cite{berkeland1998minimization}.  The sideband method applies when $\Gamma \ll \Omega_\mathrm{rf}$, and typically makes use of either a clock transition or the resolution provided by a pair of Raman beams driving a qubit transition.  In either case, presence of EMM enables the transition to be driven at the rf sideband with a coupling strength proportional to the amplitude of the EMM.   The main limitation of the sideband method is the observable frequency resolution, which is limited by laser power and decoherence.   As the micromotion amplitude is suppressed, it becomes increasingly difficult to drive the transition and maintain resonance.

Here we propose and demonstrate phase modulated sideband spectroscopy (PMSS) in which the probe laser is modulated at the trap drive frequency.  This captures the advantages of photon correlation spectroscopy and mitigates the limitations of sideband spectroscopy (SS).  When the laser is modulated, the coupling strength to the rf sideband depends on the phase of the modulation relative to the micromotion.  This leads to a differential signal proportional to the amount of EMM, which can be tuned to zero while maintaining coupling to the sideband.  The method is attractive as: (i) it can be easily implemented as a servo, (ii) it is robust against laser intensity variations, (iii) it provides enhanced sensitivity through averaging without pushing the limits of laser frequency resolution, and (iv) it can distinguish between the different types of micromotion, which appear in different quadratures.

Sideband spectroscopy with a modulated probe can be treated in the same way as conventional SS as it just adds an additional modulation to that arising from micromotion \cite{berkeland1998minimization,SM}.  Neglecting thermal effects, the coupling for a laser tuned to the micromotion sideband and modulated at the rf frequency $\Omega_\mathrm{rf}$ is given by:
\begin{equation}
\label{eq:coupling}
\Omega = \frac{\Omega_0}{2} \left[ i \left(\beta_L e^{i \phi} + \beta_m\right)-\beta_p \right]
\end{equation}
where $\Omega_0$ is a scale factor for the coupling strength of the laser, $\beta_m$ is the modulation depth from excess micromotion due displacement from the rf null, $\beta_p$ is the modulation depth due to driven motion from a phase difference $\phi_\mathrm{ac}$ between voltages applied to the rf electrodes, and $\beta_L$ is the modulation parameter for the laser, which has a phase $\phi$ relative to the trap drive.  Neglecting the rf phase-imbalance term ($\beta_p$), Rabi flopping reduces to
\begin{equation}
\label{eq:basic}
P(\Theta,R,\phi) = \frac{1}{2}\left[1+\cos \left(\Theta\sqrt{1+2 R \cos\phi+R^2} \right)\right]
\end{equation}
where $\Theta = \Omega_0 t \beta_L/2$ and $R=\beta_m/\beta_L$. For $\Theta\approx\pi/2$ and sufficiently small $R$, the population varies approximately sinusoidally with maximums and minimums at $\phi=0,\pi$.  The amplitude of the oscillation is a measure of $R$ and the midpoint a measure of $\Theta$.  

\begin{figure}
\begin{center}
\includegraphics[width=1.0 \linewidth]{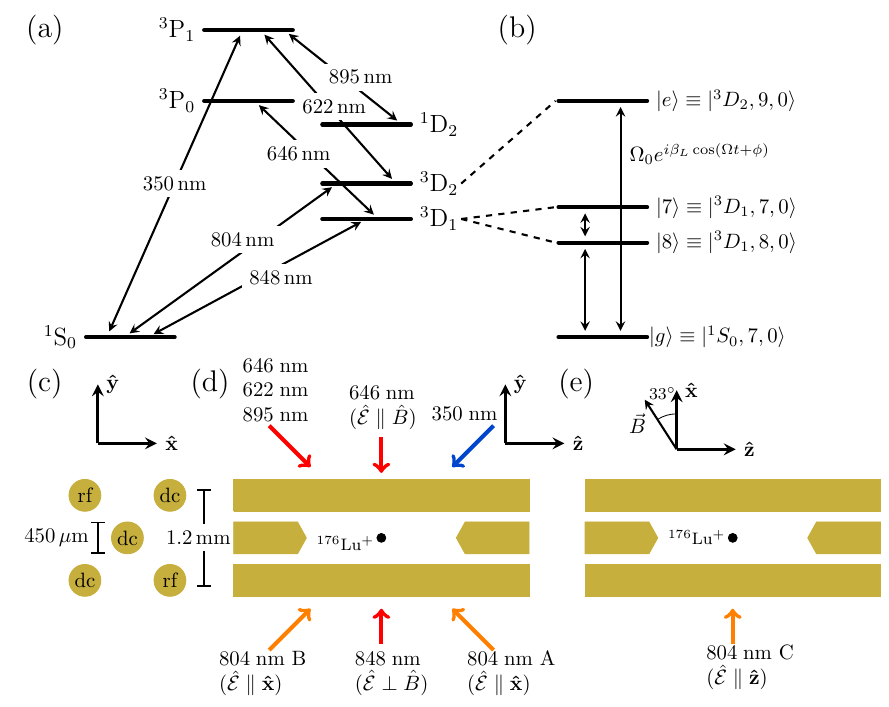}
\caption{Level structure and trap geometry used in this work.}
\label{fig:setup}
\end{center}
\end{figure}

The features of PMSS are demonstrated in two separate traps denoted Lu-1 and Lu-2 which have a similar design to those reported elsewhere \cite{kaewuam2020precision,arnold2020precision,zhiqiang2023176lu+}.  Lu-2 is the same trap used in \cite{zhiqiang2023176lu+}, but Lu-1 has been replaced to fix the heating rate reported in that work.  The relevant geometry and level structure is shown in Fig.~\ref{fig:setup}.  In a typical experiment, the atom is Doppler cooled on the 646\,nm transition, optically pumped to $\ket{^3D_1, F=7,m_F=0}$, transferred to $\ket{^3D_1, F=8,m_F=0}$ with a microwave pulse, and prepared in $\ket{^1S_0, 7,0}$ with a clock pulse on the 848\,nm transition.  Micromotion is then probed on the $\ket{^1S_0, 7,0}\leftrightarrow\ket{^3D_2, 9,0}$ transition driven on the rf sideband.  The laser modulation parameter $\beta_L$ is calibrated using heterodyne detection to determine the relative strength of carrier and sideband.  Compensation voltages are calibrated to induce a specific displacement and hence $\delta \beta_m^{(j)}$ in each of three orthogonal directions $j \in (A,B,C)$.  This is achieved by comparing coupling to the micromotion-induced sideband with an unmodulated laser, to that from a calibrated laser modulation with compensated micromotion.  The latter is modulated at a frequency slightly detuned from the trap drive to avoid any unwanted residual coupling.  

\begin{figure}
\begin{center}
\includegraphics[width=1.0 \linewidth]{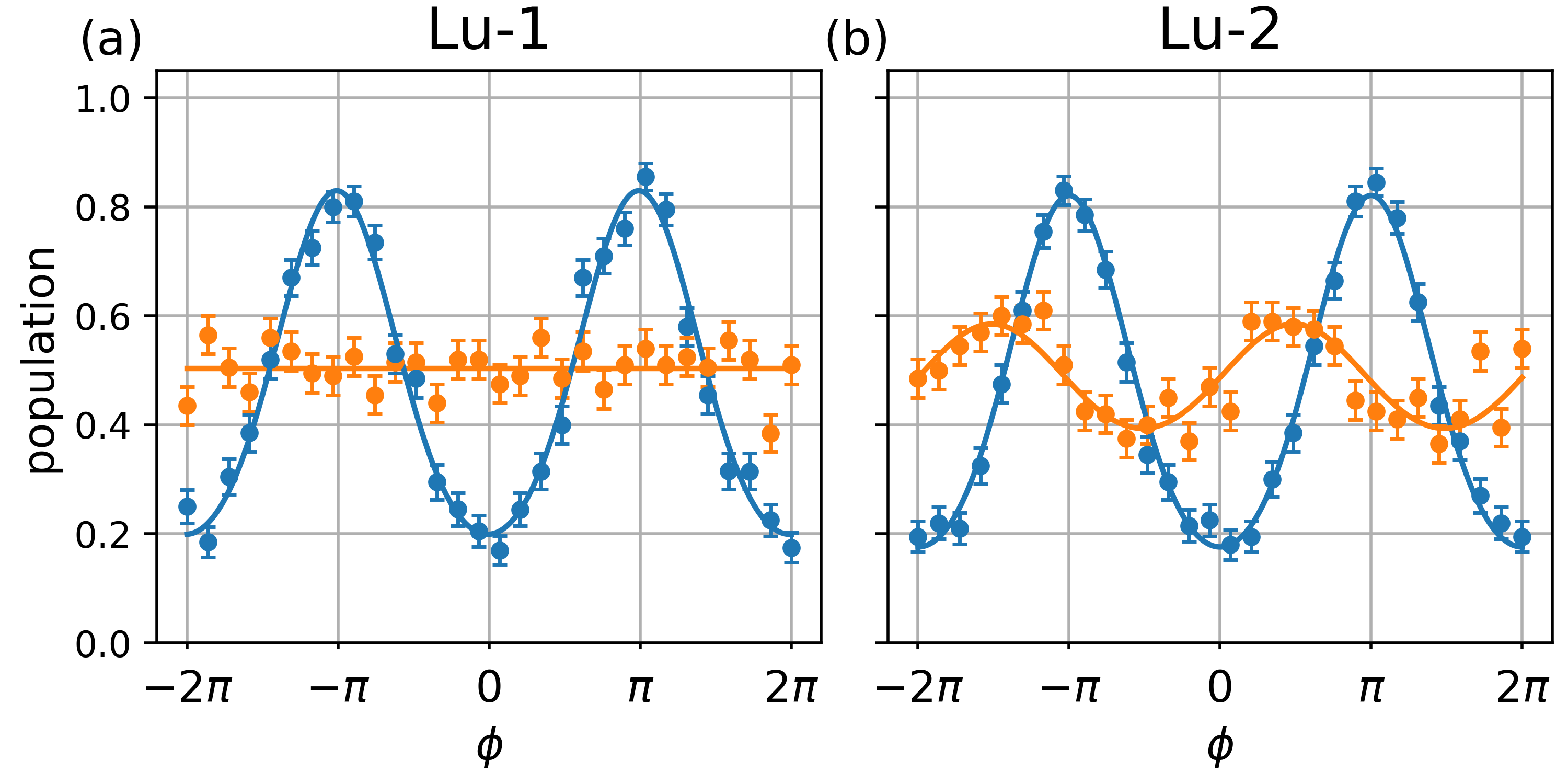}
\caption{Sideband signal as a function of the laser modulation phase.  In both curves the laser modulation is set to $\beta_L=0.1$. The blue curve corresponds to uncompensated micromotion with $\beta_m\approx 0.05$.  The orange curve is after micromotion is optimally compensated.  The residual out-of-phase oscillation for Lu-2 is due to a phase difference $\phi_\mathrm{rf}$ between voltages on the rf-electrodes.  The observed $\beta_p = 0.012$ corresponds to a $1.5\times 10^{-19}$ fractional SODS of the clock transition and a phase  $\phi_\mathrm{rf}\approx 0.3\,\mathrm{mrad}$.}
\label{fig:phase}
\end{center}
\end{figure}

We first illustrate the basic features of the method in Fig.~\ref{fig:phase}, which compares results from two traps.  In both cases the laser modulation is set to $\beta_L = 0.1$ and we scan the applied modulation phase $\phi$ after flopping for a time corresponding to $\Theta = \pi/2$.  For Lu-1, when EMM is optimally compensated (orange) there is no statistically significant variation with the phase.  However, when the ion is displaced to deliberately induce EMM with $\beta_m \approx 0.05$ (blue) we see a good fit to \eref{eq:basic} with $R=0.45$. This fit is how we determine the zero of the modulation phase for a given sign of the displacement. In Lu-2, there is clear evidence of a phase imbalance with $\beta_p = 0.012$ resulting in the phase shifted signal (orange, right side) even at optimal compensation of EMM.

\begin{figure}[t]
\begin{center}
\includegraphics[width=1.0 \linewidth]{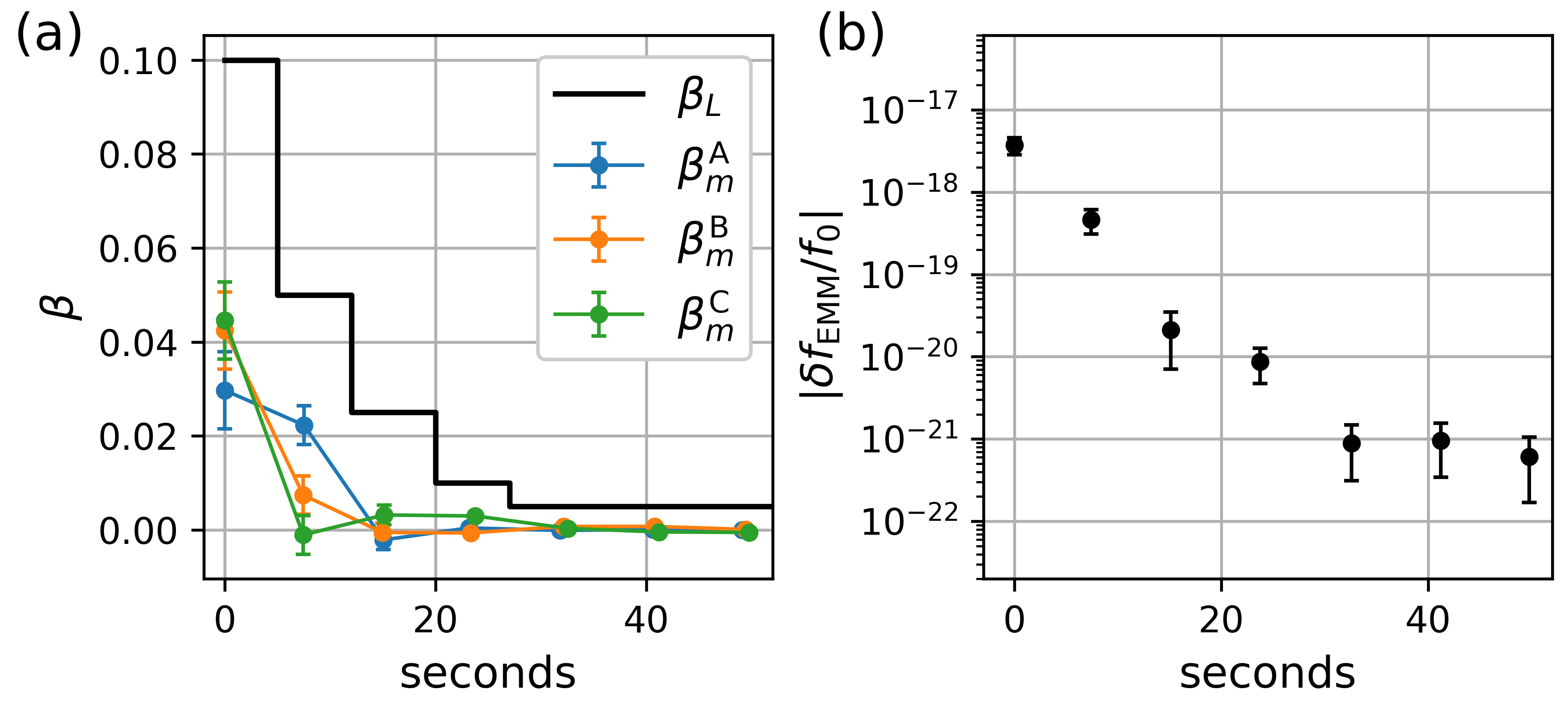}
\caption{(a) Operation of the servo starting with relatively large micromotion.  At each servo step, the value of $\beta_L$ is stepped down to enhance sensitivity to avoid excessive averaging, (b) Contribution to the fractional SODS estimated from the measurements in (a).}
\label{fig:servo}
\end{center}
\end{figure}

The EMM associated with $\beta_m\neq 0$ can be servoed to zero by using the difference in measured populations at $\phi=0$ and $\pi$ to infer $R$ and make the appropriate calibrated step change in $\beta_m$.  The mean value of the measurements $\phi=0,\pi$ also enables monitoring of $\Theta$ and hence drifts in laser intensity.  The projection-noise limit in the determination of $R$ is given by $1/(\Theta \sqrt{N})$, where $N$ is the total number of measurements i.e. $N/2$ at $\phi=0$ and $\pi$.  To avoid excessive averaging and quicken the process, $\beta_L$ can be reduced at fixed $\Theta$ to increase the sensitivity to EMM in the initial phases of the servo, which is illustrated in Fig.~\ref{fig:servo}(a). Starting with EMM deliberately uncompensated such that all $\beta_m^{(j)} \sim 0.05$, the laser modulation depth $\beta_L$ (black line) is stepped down at each servo update to increase the sensitivity to $\beta_m^{j}$ in the next measurement.  As shown in Fig.~\ref{fig:servo}(b), each servo update gains a factor of ten improvement in the associated fractional second-order Doppler shift (SODS).  Errors bars are from the projection noise in the measurement of $\beta_m^{(j)}$, taking into account the statistics of a sum of squares \cite{SM}.  When changing $\beta_L$ it is important to avoid $R>1$.  For sufficiently large $R$, the population difference can flip sign and destabilize the servo.  However, in practice, micromotion would ideally be maintained at low value with $\beta_L$ fixed at a minimum practical operating value.  If limited by laser power, $\Theta$ could also be increased to $(2k+1)\pi/2$, which is equivalent to operating at $\beta_L/(2k+1)$, although thermal dephasing and intensity variations may become a factor.
 
\begin{figure}
\begin{center}
\includegraphics[width=1.0 \linewidth]{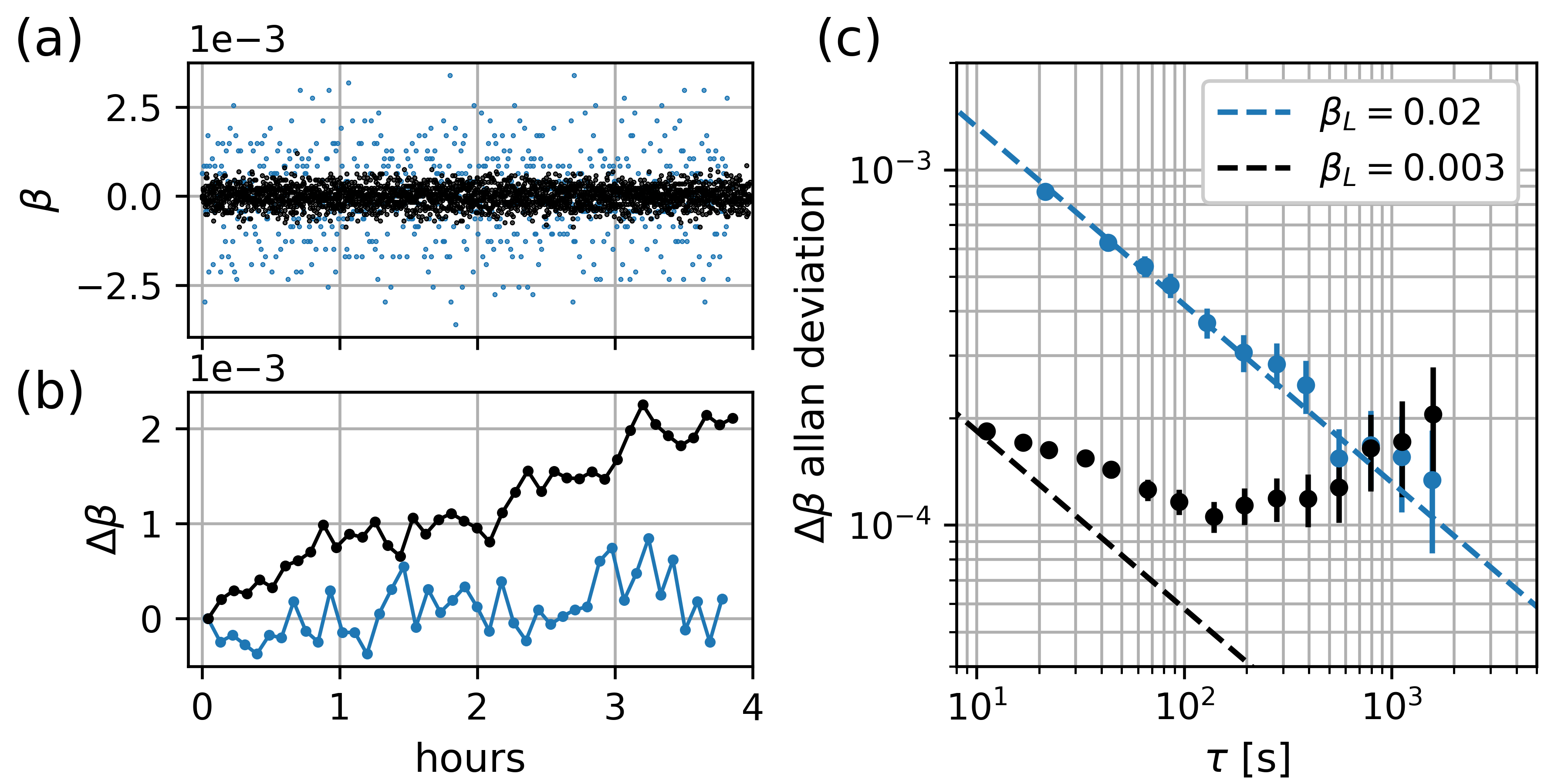}
\caption{Continuous operation of the servo showing (a)  the measured $\beta_m$ at each step, (b) the change in the compensation voltages scaled to a EMM modulation depth averaged over 5 minute intervals, and (c) the Allan deviation}
\label{fig:tracking}
\end{center}
\end{figure}

To test the limits of the method, we use all available optical power along the radial direction (denoted C in Fig.~\ref{fig:setup}) to access smaller values of $\beta_L$.  For Lu-1, we can obtain a carrier coupling of $\Omega_0 \approx 2\pi \times 15\,\mathrm{kHz}$, which enables us to operate the servo at a modulation depth of $\beta_L=0.003$.  This corresponds to a sideband $\pi/2$-time of $\sim$11\,ms, which is well within the coherence time of the laser. Fig.~\ref{fig:tracking} demonstrates continuous operation of the servo for around four hours at two values of $\beta_L$.  Fig.~\ref{fig:tracking}(a) shows the measured $\beta_m$ at each servo update, and Fig.~\ref{fig:tracking}(b) the change in the compensation voltages scaled to a EMM modulation depth in that direction relative to the starting point.  The latter are averaged over 5 minute intervals for clarity.  The Allan deviation for each data set is given in Fig.~\ref{fig:tracking}(c).  For $\beta_L=0.02$, we are limited by averaging time, but at the more extreme $\beta_L=0.003$, we become limited by drifts in the associated stray fields such that averaging longer than a few minutes has no benefit.  For longer times we see that without servo tracking, the EMM would drift out of compensation on the order of $\delta \beta \sim 1\times10^{-3}$ in one hour, which corresponds to a fractional clock shift of $7\times10^{-22}$. Given the sensitivity of the measurement and the relative passive stability of the EMM, we can expect to maintain EMM below $10^{-20}$ with short ($< 1$ minute) automated measurements every few hours using this technique. 

Measurement time includes the overhead of cooling, state preparation, and maintaining resonance of the 848\,nm laser with the clock transition.  During normal clock operation, the atom would be registered in the ground state approximately 50\% of the time.  Given the observed stability of EMM, a more optimal strategy would be to accumulate EMM measurements each time a clock interrogation registers the atom in the ground state.  Assuming an interrogation time of approximately 1\,s, we would expect to accumulate approximately 600 measurements in 20 minutes with minimal disruption to clock operation.  Operating at $\beta_L=0.01$ would then provide a projection-limited $\delta \beta_m$ of $5\times 10^{-4}$ in all three directions, which is sufficient to compensate the observed drift and maintain a mean clock shift of $5\times 10^{-22}$ from EMM.

Conventional SS has a thermally limited $\beta_\mathrm{min}$, which may be written $\beta_\mathrm{min}\approx q\eta^2 (\bar{n}+\tfrac{1}{2})$, where $\eta=k\sqrt{\hbar/(2m\omega^2)}$ is the Lamb-Dicke parameter and $q$ the usual Mathieu parameter associated with the rf trapping potential \cite{meir2018experimental,keller2015precise}.  For our typical parameters $\beta_\mathrm{min}\approx 10^{-3}$, so we are able to operate well below this limit.  To properly interpret the behaviour in this regime we must modify Eq.~\ref{eq:coupling} to account for the thermal coupling and that we probe at $45^\circ$ to the principle axes of the rf potential.  This adds an additional term $\tfrac{1}{2}\Omega_0\left[\alpha_2 \left(n_2+\tfrac{1}{2}\right)-\alpha_1\left(n_1+\tfrac{1}{2}\right)\right]$ to Eq.~\ref{eq:coupling}, where $\alpha_i=q_i \eta_i^2$.  The resulting change to Eq.~\ref{eq:basic} must then be averaged over the thermal distributions.  

When $\alpha_i=\alpha$ and $\bar{n}_i=\bar{n}$, there are two unique signatures of intrinsic micromotion when driving the rf-sideband.  Firstly, when driving with unmodulated light, there is no longer any Rabi flopping as such.  Instead, population is driven to $0.5$ but revives in a time $T_r=4\pi/(\Omega_0 \alpha)$.  For times $t\ll T_r$,  
\begin{equation}
P(t)\approx \frac{1}{2}\left(1+\frac{1}{1+(\gamma t/)^2}\right),
\end{equation} 
where $\gamma=\tfrac{1}{2}\Omega_0 \alpha \sqrt{\bar{n}(\bar{n}+1)}$.  Secondly, when driven with modulated light, the dependence on the modulation phase $\phi$ becomes $\pi$-periodic with maximums of 0.5 at $\pm\pi/2, \pm 3\pi/2$ when $\Theta=\pi/2$.  Small asymmetries in parameters do not significantly change these conclusions, particularly at larger $\bar{n}$. 
\begin{figure*}
\begin{center}
\includegraphics[width=1.0 \linewidth]{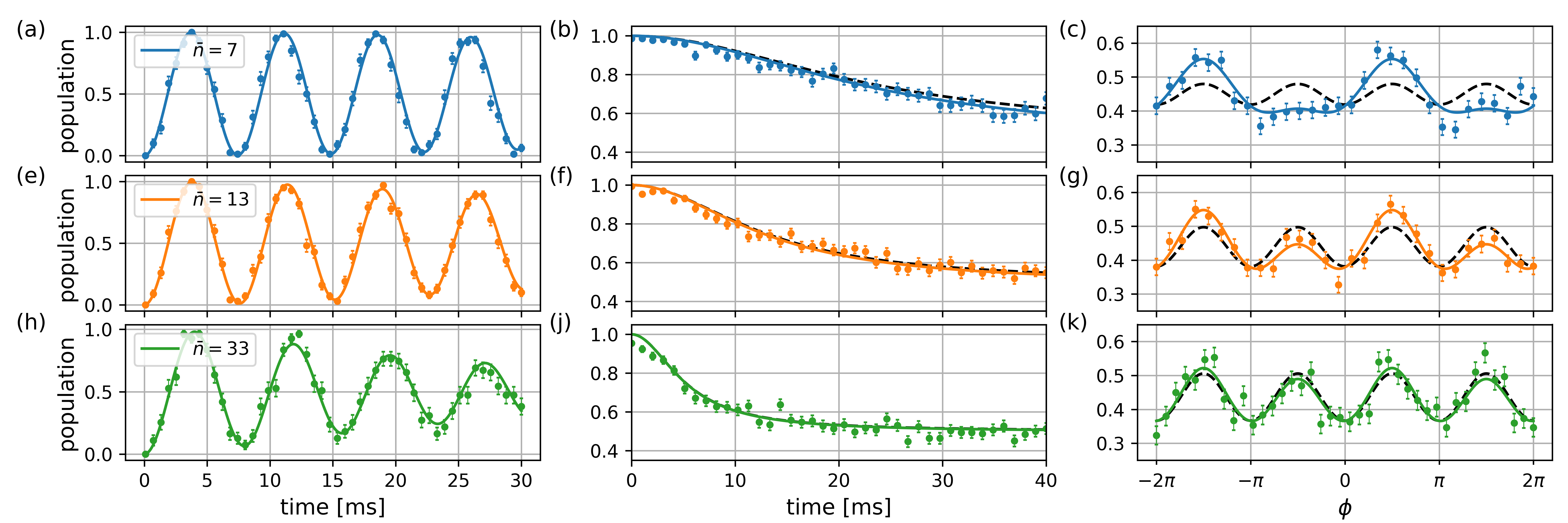}
\caption{Behaviour at the thermal limit.  Left column: thermal dephasing when flopping on the 848\,nm transition, from which we infer $\bar{n}$. Middle column: driving the rf sideband on the 804\,nm transition with unmodulated light.  Right column: scan of the phase when driving on the rf sideband with modulated light ($\beta_L=3\times10^{-3}$, $\Theta=\pi/2$).  For center and right plots, dashed curves are ab initio calculations with $\beta_p=\beta_m=0$, solid curves include $\beta_p = -3.1\times10^{-4}$, which corresponds to  $\phi_\mathrm{rf}\approx 5\,\mathrm{\mu rad}$.  For these experiments $\Omega_\mathrm{rf}=2\pi\times 20.7\,\mathrm{MHz}$ and $(\omega_1,\omega_2) = 2\pi\times(615,650)\,\mathrm{kHz}$ giving $(\eta_1,\eta_2)=(0.038,0.037)$.  We assume $\bar{n}_1=\bar{n}_2$ and the approximation $q_i\approx \omega_i \sqrt{8}/\Omega_\mathrm{rf}$ gives $\alpha_1=\alpha_2=1.2\times 10^{-4}$.}
\label{fig:thermal}
\end{center}
\end{figure*}

In Fig.~\ref{fig:thermal}, we illustrate the behaviour at different $\bar{n}$, which is controlled by varying the cooling beam detuning.  The left most column shows the thermal dephasing when flopping on the 848\,nm carrier from which we infer $\bar{n} = (7,12,33)$ for the upper, middle, and lower plots respectively.   The middle column shows flopping on the micromotion sideband with no modulation ($\beta_L =0$) and EMM compensated ($\beta_m < 2\times10^{-4}$).  The right column shows a scan of the applied modulation phase $\phi$ with $\beta_L=3\times10^{-3}$ and the probe time set such that $\Theta=\pi/2$. For both center and right plots, the dashed lines are ab initio calculations using only thermal coupling ($\beta_m,\beta_p =0$). The solid lines include $\beta_p = -3.1\times10^{-4}$ which was optimised as a free parameter to account for the observed asymmetry and corresponds to a phase  $\phi_\mathrm{rf}\approx 5\,\mathrm{\mu rad}$.  Accurate assessment of $\beta_p$ would require a more careful characterization of trap parameters to determine the contribution from inherent trap asymmetry but this alone cannot account for the observations.

Comparison of different micromotion compensation methods is complicated by the constraints of any particular experiment and the properties of the atom used.  However, PMSS is a readily implemented enhancement over the widely used SS.  Whatever minimum $\beta_m$ can be achieved with SS defines the minimum $\beta_L$ that can be used for PMSS, from which one can continue to average down if required.  In our case we have reached a limit of $\delta \beta_m=1.35\times10^{-4}$ corresponding to: a $1.4\times 10^{-23}$ contribution to the fractional SODS; compensation of stray electric fields to $0.01\,\mathrm{Vm^{-1}}$ comparable to that obtained using neutral atoms as a probe \cite{harter2013minimization}; and a displacement of the ion relative to the rf null by 0.4\,nm, which is approximately 5\% of the rms size of the ground-state wavefuntion.  Our approach has also enabled us to monitor the drifts of stray electric fields over the timescales of five minutes.  The observed changes correspond to drift velocities of a few nm/hr with associated fractional first order Doppler shifts of low $10^{-21}$.

Motion is a fundamental consideration in the performance of an ion-based optical atomic clock.  It is quantum mechanically unavoidable and intrinsic micromotion an inescapable consequence of confinement in a Paul trap.  For Lu-1, the ground state energy is associated with a SODS of $\approx 2\times 10^{-20}$.  We have demonstrated the ability to suppress all other excess motion to well below this fundamental limit in a way that can be seamlessly integrated into clock operation. This has been made possible through the use of PMSS together with the high performance secondary clock transition that lutetium uniquely provides.

This research is supported by the Agency for Science, Technology and Research (A*STAR) under Project No. C210917001; the National Research Foundation (NRF), Singapore, under its Quantum Engineering Programme (QEP-P5); the National Research Foundation, Singapore and A*STAR under its Quantum Engineering Programme (NRF2021-QEP2-01-P03) and its CQT Bridging Grant; and the Ministry of Education, Singapore under its Academic Research Fund Tier 2 (MOE-T2EP50120-0014).
\bibliography{EMM}
\newpage\clearpage


\section*{Supplementary material}

# Supplemental: Enhanced micromotion compensation using a phase modulated light field.

K. J. Arnold, N. Jayjong, M. L. D. Kang, Qin Qichen, Zhao Zhang, and Qi Zhao
*Centre for Quantum Technologies, National University of Singapore, 3 Science Drive 2, 117543 Singapore*

M. D. Barrett
*Centre for Quantum Technologies, National University of Singapore, 3 Science Drive 2, 117543 Singapore*
*Department of Physics, National University of Singapore, 2 Science Drive 3, 117551 Singapore and*
*National Metrology Center, Agency for Science, Technology and Research (A*STAR), Singapore*

We give the derivations of equations for coupling a phase-modulated laser to an atom via the rf sideband and a few analytic results when the coupling is dominated by intrinsic micromotion. We also briefly comment on the proper statistical evaluation of a sum of squares, which is relevant to the evaluation of the second order Doppler shift from measured $\beta$-parameters.

## COUPLING EQUATIONS

To illustrate the salient features of our method, we first consider only motion and probing along one dimension and follow the methodology given in [1]. The lowest order solution to the Mathieu equation is given by

$$x = (u_0 + u\cos(\omega t + \phi_s))\left(1 + \tfrac{q}{2}\cos\Omega_{\rm rf}t\right), \quad (S1)$$

where $u_0$ is a displacement due to uncompensated stray electric fields, $\Omega_{\rm rf}$ is the trap drive radio-frequency (rf), $u$ is the amplitude of motion at the pseudo-potential trapping frequency $\omega$ and $q$ is the usual parameter from the Mathieu equation related to the applied rf-voltage. Taking this motion to be along the axis of a probe laser results in an effective frequency modulation of the laser, which may be expressed

$$E_0 e^{i(\omega_L t + \beta_L \cos\theta_L + \beta_m \cos\theta_m + \beta_s \cos(\theta_s) + \beta\cos\theta_+ + \beta\cos\theta_-)}$$

where $\beta_L$ and $\theta_L = \Omega_{\rm rf}t + \phi$ describes the modulation placed on the light-field, $\beta_m = ku_0 q/2$ and $\theta_m = \Omega_{\rm rf}t$ describes the modulation arising from excess micromotion, $\beta_s = ku$, $\theta_s = \omega t + \phi_s$, $\beta = \tfrac{1}{4}kqu$ and $\theta_\pm = \Omega_{\rm rf}t \pm \phi_s$ describes the modulation from the intrinsic motion of the ion. To lowest order, coupling at the rf sideband is then given by

$$\frac{\Omega_0}{2}\left[i(\beta_L \cos\phi + \beta_m) - (\beta_L \sin\phi + \beta_T)\right] \quad (S2)$$

where $\beta_T = \tfrac{1}{4}q\beta_s^2$ is the modulation associated with intrinsic micromotion from the secular motion of the ion. This does not include excess micromotion associated with the phase difference between rf electrodes. In the classical approximation, $u$ is the amplitude of the secular motion so we use $m\omega^2 u^2/2 = E = \hbar\omega(n+\tfrac{1}{2})$, giving

$$\beta_T = \frac{1}{4}q\beta_s^2 = q\eta^2(n+\tfrac{1}{2}), \quad (S3)$$

where $\eta = k\sqrt{\hbar/(2m\omega)}$ is the Lamb-Dicke parameter. Results are then averaged over the thermal distribution.

A full quantum treatment of the Mathieu equation is given in [2, 3]. Following that work, the atom-laser interaction is written

$$\frac{\hbar\Omega_0}{2}\sigma^+ e^{i\eta(au^*(t)+a^\dagger u(t))-i\delta t+i\beta_L \cos\theta_L + i\beta_m \cos\theta_m} + h.c.$$

where

$$u(t) \approx e^{i\omega t}\left(1 + \tfrac{q}{2}\cos\Omega_{\rm rf}t\right),$$

and we have treated the excess micromotion term as an additional modulation on the laser. In calculating the coupling to vibrational states, we use

$$e^{-|\eta u(t)|^2/2}L_n^0(|\eta u(t)|^2) \approx \left(1 - \left(n+\tfrac{1}{2}\right)\eta^2|u(t)|^2\right).$$

and

$$|u(t)|^2 \approx 1 + \tfrac{1}{2}q\left(e^{i\Omega_{\rm rf}t} + e^{-i\Omega_{\rm rf}t}\right).$$

To lowest order, coupling at the rf sideband is then given by

$$\frac{\Omega_0}{2}\left[i\left(\beta_L e^{i\phi} + \beta_m\right)\left(1 - \eta^2\left(n+\tfrac{1}{2}\right)\right) - \eta^2 q\left(n+\tfrac{1}{2}\right)\right].$$

Typically, $\beta_L, \beta_m \ll q$, and the coupling is then

$$\frac{\Omega_0}{2}\left[i(\beta_L \cos\phi + \beta_m) - \left(\beta_L \sin\phi + q\eta^2\left(n+\tfrac{1}{2}\right)\right)\right].$$

exactly as before. A phase imbalance on the electrodes may also drive excess micromotion that is out of phase with the trap drive, which has not been included. This adds an additional modulation term that is shifted by $\pi/2$ relative to $\beta_m$ and can be included with the replacement $\beta_m \to \beta_m + i\beta_p$.

When the probe laser propagates at $45°$ to the principal axes, the only change is to the values of $q$ and $\eta$ and we make the replacement

$$q\eta^2\left(n+\tfrac{1}{2}\right) \to \alpha_1^2\left(n_1+\tfrac{1}{2}\right) - \alpha_2^2\left(n_2+\tfrac{1}{2}\right). \quad (S4)$$

To a good approximation $\alpha_1 \approx \alpha_2 = \alpha$.

## PROJECTION NOISE

Starting with
$$P(\Theta, R, \phi) = \tfrac{1}{2}\left(1 + \cos\left[\Theta\sqrt{1 + 2R\cos\phi + R^2}\right]\right),$$
we have the differential population given by
$$S(\Theta, R) = P(\Theta, R, \pi) - P(\Theta, R, 0)$$
$$= \sin\Theta \sin(\Theta R) \quad\text{(S5)}$$
with variance
$$\text{Var}(\Theta, R) = \tfrac{1}{2}\left(1 - \cos\Theta\cos(\Theta R)\right). \quad\text{(S6)}$$
For the projection noise limit, we are primarily interested in small $R$ for which we have
$$\frac{\sqrt{\text{Var}(\Theta, R)}}{\partial S/\partial R} \approx \frac{1}{\Theta\sqrt{2}} + \mathcal{O}(R^2), \quad\text{(S7)}$$
and the approximation is exact when $\Theta = \pi/2$. So for a total of $N$ measurements i.e. $N/2$ at both $\phi = 0, \pi$, we have $1/(\Theta\sqrt{N})$. We also have the mean population
$$\tfrac{1}{2}(P(\Theta, R, \pi) + P(\Theta, R, 0)) = \tfrac{1}{2}\left(1 + \cos\Theta\cos(\Theta R)\right).$$
For small $R$,
$$S(\Theta, R) \approx (\Theta \sin\Theta) R.$$
Consequently, $\Theta \approx 0.65\pi$ makes the effective gain most robust to small changes in $\Theta$.

## POPULATIONS

When excess micromotion is completely absent and the laser unmodulated, only the intrinsic micromotion can provide a coupling, which may be evaluated using the identity
$$\frac{1}{1+\bar{n}}\sum_{n=0}^{\infty}\left(\frac{\bar{n}}{1+\bar{n}}\right)^n e^{in\theta} = \frac{1}{\bar{n}+1-\bar{n}e^{i\theta}}. \quad\text{(S8)}$$

Assuming both modes have the same $\alpha$ and $\bar{n}$, we have
$$P(\theta) = \frac{1}{2}\left(1 + \frac{1}{1 + 4\bar{n}(\bar{n}+1)\sin^2\left(\frac{\theta}{2}\right)}\right) \quad\text{(S9)}$$
where $\theta = \tfrac{1}{2}\Omega_0\alpha t$. For typical parameters, population is quickly driven to 0.5 but revives in a time $T_R = 4\pi/(\Omega_0\alpha)$. For our parameters, $T_R \approx 1\,\text{s}$ and beyond the coherence time of our laser. Moreover, at such long timescales higher order terms may become significant. For times $t \ll T_R$,
$$P(t) \approx \frac{1}{2}\left(1 + \frac{1}{1 + (\gamma t/)^2}\right), \quad\text{(S10)}$$
where $\gamma = \tfrac{1}{2}\Omega_0\alpha\sqrt{\bar{n}(\bar{n}+1)}$. In this approximation, there is no coupling at $\bar{n} = 0$. For smaller $\bar{n}$, one would need to account for the differences in $\alpha_i$, and $\bar{n}_i$.

When the laser is modulated, population for a given $n_1$ and $n_2$ is given by
$$\frac{1}{2}\left[1 + \cos\left(\Theta\sqrt{1 + 2R_{n_1,n_2}\sin\phi + R_{n_1,n_2}^2}\right)\right],$$
where $\Theta = \Omega_0 t \beta_L/2$ and
$$R_{n_1,n_2} = \bar{\alpha}_1(n_1 + \tfrac{1}{2}) - \bar{\alpha}_2(n_2 + \tfrac{1}{2}),$$
with $\bar{\alpha}_i = q_i\eta_i^2/\beta_L$. When $\bar{\alpha}_1 = \bar{\alpha}_2$ and the thermal distributions are the same in both directions, a change in $\phi$ by $\pi$ is equivalent to an interchange of dummy summation variables in the thermal average. Hence the population is $\pi$-periodic in $\phi$ under isotropic conditions. One can also use Eq. S8 to show that the populations $\phi = \pm\pi/2, \pm 3\pi/2$ are 0.5 when $\Theta = \pi/2$. Thus the population has a double oscillation with the mean value displaced from 0.5 by an amount that increases with $\bar{n}$. The degree that the symmetry is broken when the system is not strictly isotropic can be determined by writing $\alpha_i = \bar{\alpha} \pm \tfrac{1}{2}\delta\bar{\alpha}$, $\bar{n}_i = \bar{n} \pm \tfrac{1}{2}\delta\bar{n}$ and determining the population difference between $\phi = \pm\pi/2$. To lowest order in $\delta\bar{n}$ and $\delta\alpha$, we find
$$\frac{\left(\bar{n}+\tfrac{1}{2}\right)\delta\bar{\alpha} + \delta\bar{n}\sin\bar{\alpha}}{\left[\cos^2\tfrac{\bar{\alpha}}{2} + (2\bar{n}+1)^2\sin^2\tfrac{\bar{\alpha}}{2}\right]^2}\sin\Theta.$$
A phase shift between rf-electrodes adds $\beta_p/\beta_L$ to $R_{n_1,n_2}$ and dominates the behaviour when $\beta_p/\beta_L \gtrsim \bar{\alpha}\bar{n}$.

## SHIFTS AND UNCERTAINTIES

Since $\beta_m^{(j)}$ is servoed to zero, we consider the distribution of $\beta^2 = \sum_j \left[\beta_m^{(j)}\right]^2$, which is relevant to the interpretation of the second order Doppler shift. In general, if $n$ random variables $X_k$ are $\mathcal{N}(\mu_k, 1)$ i.e. normally distributed with mean values $\mu_k$ and unit variances, the sum of squares is determined by $\mathcal{Z}(n, \lambda)$: a non-central chi-squared distribution with $n$ degrees of freedom, with noncentrality parameter $\lambda = \sum_k \mu_k^2$. This distribution has a mean $3 + \lambda$ and variance $6 + 4\lambda$. The probability of being within $\pm\sqrt{6 + 4\lambda}$ of the mean, limits to 0.683 at large $\lambda$ as expected for a normal distribution, but limits to 0.766 at $\lambda = 0$ as for a chi-square distribution with 3 degrees of freedom. As a better fit to the notion of a $1\sigma$ uncertainty, we take $\sigma = \sqrt{a + 4\lambda}$, where $a = 4.711$. The probability of being within $\pm\sigma$ of the mean agrees with that for a normal distribution at $\lambda = 0$ and $\lambda \gg 1$, deviates by no more than 0.02 for any $\lambda$, and is symmetric about the mean to the same level. These results can be rescaled to account for non-unity variance.

If each $\beta_m^{(j)}$ is measured with a projection-noise limit $\sigma$, the mean second order Doppler shift will be determined by

$$\left(\frac{\Omega_{\rm rf}}{2ck}\right)^2 \left(\beta^2 + 3\sigma^2 \pm \sigma\sqrt{4\beta^2 + 4.711\sigma^2}\right), \qquad (S11)$$

where $c$ is the speed of light. For $\beta \gg \sigma$, this agrees with the often quoted quadrature sum of $2\beta_m^{(j)}\sigma$. If continuously updating a servo, we would take $\beta = 0$. However, if compensation was set by a particular measurement, we would use the measured $\beta$.

The same treatment can be used for $\beta_p$. However, since we expect $\beta_p$ to be associated with motion predominately along one principle axis, we can consider $2\beta_p^2$ when measuring at the 45° angle. This is then associated with one degree of freedom. In this case we can numerically calculate the 68% confidence interval with equal probabilities about the mean.

The origin of uncompensated EMM associated with $\beta_p$ is assumed to be the result of a phase difference between voltages on the rf electrodes. With a phase difference $\phi_{\rm rf}$,

$$\beta_p = \left(\frac{k}{\sqrt{2}}\right)\left(\frac{Q}{m\Omega_{\rm rf}^2}\right)\left(\frac{V_{\rm rf}\phi_{\rm rf}\alpha}{2R}\right), \qquad (S12)$$

where $Q$ is the ion charge, $V_{\rm rf}$ is the amplitude of the trap drive, $R \approx 0.62$ mm is the minimum ion to electrode distance, and $\alpha$ is a trap dependent geometry factor on the order of unity. This expression accounts for the 45° angle of the probe to the principal axes. From simulations of trapping potentials, $\alpha \approx 0.7$. For Lu-1, $\Omega_{\rm rf} = 2\pi \times 20.7$ MHz and $V_{\rm rf} \approx 630$ V giving $\beta_p \approx 65\phi_{\rm rf}$. For Lu-1, $\Omega_{\rm rf} = 2\pi \times 16.9$ MHz and $V_{\rm rf} \approx 370$ V giving $\beta_p \approx 40\phi_{\rm rf}$.

---


\end{document}